\documentclass[a4paper,11pt]{article}
\pdfoutput=1

\usepackage[pdftex]{hyperref}
\usepackage{amsmath}
\usepackage{amssymb}
\usepackage{fullpage}
\usepackage{cancel}
\usepackage{float}
\usepackage[affil-it]{authblk}
\usepackage[hang,flushmargin]{footmisc}
\usepackage{color}

\newcommand{\ar}{$a_r^{\! (1)}\,$}

\numberwithin{equation}{section}

\hypersetup{ pdfborder = {0 0 0}}

\title{Folding defect affine Toda field theories}
\author{Craig Robertson\footnote{\href{mailto:craig.robertson@durham.ac.uk}{craig.robertson@durham.ac.uk}}}
\affil{Department of Mathematical Sciences, Durham University, Durham, DH1 3LE, UK}
\date{}

\begin{document}

\begin{titlepage}

\renewcommand{\thefootnote}{\alph{footnote}}
\clearpage\maketitle
\thispagestyle{empty}
\renewcommand{\thefootnote}{\arabic{footnote}}

\begin{abstract}
\noindent
A folding process is applied to fused $a_r^{\! (1)}$ defects to construct defects for the non-simply laced affine Toda field theories of $c_n^{(1)}$, $d_n^{(2)}$ and $a_{2n}^{\! (2)}$ at the classical level. Support for the hypothesis that these defects are integrable in the folded theories is given by the demonstration that energy and momentum are conserved. Further support is provided by the observation that transmitted solitons retain their form.
\end{abstract}
\end{titlepage}

\newpage

\section{Introduction}

The subject of affine Toda field theory (ATFT) has seen an upsurge of interest due to the discovery of integrable defects. Whilst much is known about defects in the sine-Gordon model \cite{KL,BCZ03, BCZ05, HK, CZ10a, AAGZ, AD}; the same cannot be said about the ATFTs in general. Indeed, thus far only for the \ar \cite{BCZ04} and $a_2^{\! (2)}$ \cite{CZ09b} models do integrable defects exist in the literature - even at the classical level - despite the search for defect Lagrangians in ATFT having been initiated a decade ago \cite{BCZ03}. Compare this to the discovery of solitons in ATFT, where the construction of \ar and $d_4^{(1)}$ solitons \cite{Holl} was rapidly followed by solitons in the other models \cite{MM,McGhee}; or to how the discovery of integrable boundary conditions in the sine-Gordon model \cite{GZ} was soon extended to all other ATFTs \cite{BCDR}. The different ATFTs have similar properties (e.g., they all stem from the root data of semi-simple Lie algebras - see the early references \cite{MOP,Wils,OT1,OT2}) so it is expected that defects should exist for all of the ATFTs - as such, an overarching goal in this field is to find and investigate the properties of all of the possible defects. \\
\\
Folding, or reduction, is a powerful tool which allows properties of non-simply laced theories to be found using properties of simply laced theories\footnote{The simply laced ATFTs are the $a_r^{\! (1)}$,$d_s^{(1)}$,$e_6^{(1)}$,$e_7^{(1)}$ and $e_8^{(1)}$ theories. They are distinguished in having all roots of the same length, conventionally $\sqrt{2}$; although in the $a_1$ case the choice of $\alpha=1$ is more conventional. The non-simply laced theories all have more than one length of root present.} which are often easier to work with. Indeed, folding has previously been put to use to find, in the non-simply laced theories, solitons \cite{MM,McGhee} and also scattering matrices at the classical \cite{BCDS} and quantum \cite{Khas} levels. Note that, except in the simplest case of $a_2^{\! (1)} \to a_2^{\! (2)}$ \cite{CZ09b}, folding has not previously been applied to the study of defect ATFTs.\\
\\
In this paper, a folding process, originally described in \cite{KS}, is used to obtain candidate integrable defects for the $c_n^{(1)}$, $d_n^{(2)}$ and $a_{2n}^{(2)}$ series of non-simply laced ATFTs by making use of \ar solitons and defects. Two methods, which turn out to be linked, are used to suggest that these defects should be classically integrable. Firstly, in what circumstances does the defect conserve energy and momentum; secondly, what happens when a soliton is sent through the defect. An interesting by-product of the folding process is that it allows for the construction of multisolitons, breathers and fusing rules for these folded theories ($c_n^{(1)}$, $d_n^{(2)}$ and $a_{2n}^{(2)}$) - something not explicitly considered previously in the literature. \\
\\
What this paper does not do is provide defects for any of the ATFTs which do not fall under the umbrella of \ar , though it does suggest that if the appropriate defects can be found for the other simply laced theories ($d_s^{(1)}$, $e_6^{(1)}$, $e_7^{(1)}$ and $e_8^{(1)}$) then folding may be applied to give the defects of the other non-simply laced theories. Note that only one type of defect is given for each of $c_n^{(1)}$, $d_n^{(2)}$ and $a_{2n}^{\! (2)}$ - there is no claim that other defects should not exist, and indeed there is reason to believe that there will be other defects, in these theories.
\\
\\
In order to set notation, a short summary of ATFT with defects is given below. \\
\\
To each affine Dynkin diagram there is a corresponding affine Toda model \cite{OT1,Dorey}. In 1+1 dimensions a working definition of ATFT is given by the Lagrangian
\begin{align}
\mathcal{L}(u) = \frac{1}{2}\dot{u}\cdot \dot{u} - \frac{1}{2}u' \cdot u' - U(u) \label{lag1}
\end{align}
where $u$ is an $r$ component vector living in the root space described by the affine Dynkin diagram and is a Lorentz scalar. The potential is given by \cite{Holl,MM}
\begin{align}
U(u) = \frac{m^2}{\beta^2} \sum_{j=0}^r n_j \left( e^{\beta \alpha_j \cdot u} - 1 \right) \; . \label{pot1}
\end{align}
\\
In \eqref{pot1}, $\{\alpha_i\}$ ($i=1,\ldots,r$), are the positive simple roots of the root space while $\alpha_0 = - \sum_{j=1}^r n_j \alpha_j$ is the lowest root of the root space, corresponding to the extra node of the affine Dynkin diagram. It is the case that $\sum_{j=0}^r n_j \alpha_j = 0$, so conventionally $n_0=1$, while the other marks $\{ n_i \}$ are characteristic of the underlying Lie algebra. The constant $m$ sets a mass scale, which has no importance for the classical discussions herein so will be set to unity, $m=1$. The coupling constant is denoted $\beta$. The potential here is chosen such that the zero solution has zero energy. \\
\\
Guided by the Lagrangian description of a sine-Gordon defect \cite{BCZ03}; Bowcock, Corrigan and Zambon suggested an ansatz for defects in the other ATFTs in \cite{BCZ04} (subsequently referred to as a type I defect \cite{CZ09b})
\begin{align}
\mathcal{L}=\theta(-x)\mathcal{L}_u + \theta(x)\mathcal{L}_v + \delta(x)\left(\frac{1}{2}uA\dot{u} + uB\dot{v} + \frac{1}{2}vC\dot{v} - D(u,v)\right) \; . \label{lag2} 
\end{align}
\\
Equation \eqref{lag2} describes a defect situated at $x=0$ with $A$, $B$ and $C$ constant matrices while $\mathcal{L}_u$ and $\mathcal{L}_v$ are `bulk' Lagrangians of the form \eqref{lag1} for the fields $u$ and $v$ respectively. In practice $u$ and $v$ always correspond to the same root data. \\
\\
The Euler--Lagrange equations applied to \eqref{lag2} give at $x=0$ the conditions
\begin{align}
u' &= A\dot{u} + B\dot{v} - D_u  \nonumber \\
v' &= -C\dot{v} + B^T\dot{u} + D_v \label{defcon1}
\end{align}
as well as the bulk equations of motion. The notation $D_u$ means the gradient of $D$ with respect to the vector field $u$. \\
\\
In \cite{BCZ04} a defect Lax pair, taking into account \eqref{defcon1}, was constructed and it was shown that only the \ar models may incorporate such a defect and remain integrable. It was subsequently shown in \cite{CZ09a} that \eqref{lag2} conserves modified energy and momentum only for \ar and that the conditions imposed are the same as for the Lax pair - i.e., in this case energy-momentum conservation implies classical integrability. From either approach it is found that
\begin{align}
A&=C=1-B \nonumber \\
D(u,v) &= d \sum_{j=0}^r e^{\stackrel{\frac{1}{2}\alpha_j \cdot \left( B^T u + Bv \right)}{}} + d^{-1} \sum_{j=0}^r e^{\stackrel{\frac{1}{2}\alpha_j \cdot B\left( u - v \right)}{}} \label{defint}
\end{align}
and so the defect can be classified by $B$ along with the parameter $d$. It was noted in \cite{BCZ04} that $d=e^{-\eta}$ with $\eta$ behaving like a rapidity associated with the defect, so one can say that the terms with a prefactor $d$ have negative helicity, while those with $d^{-1}$ have positive helicity. Of relevance to this paper, there are two possibilities for $B$, which are
\begin{align}
B &= \sum_{j=1}^r \left(\lambda_j - \lambda_{j+1}\right)\lambda_j^T  \label{bform1}
\end{align}
and its transpose
\begin{align}
B &= \sum_{j=1}^r \left(\lambda_j - \lambda_{j-1}\right)\lambda_j^T  \label{bform2}
\end{align}
where $\{\lambda_i\}$ are the fundamental highest weights of \ar, defined by $\lambda_i \cdot \alpha_j = \delta_{ij}$ for $i,j=1,\ldots,r$ and $\lambda_0 = 0$. In this paper $B$ will be taken to mean \eqref{bform1}, with $B^T$ used if the defect is classified by \eqref{bform2}.
\\
\\
In \cite{BCZ04,CZ07} the effect of a defect on a one soliton solution was examined. Using a one soliton ansatz of species $p$~\footnote{For \ar solitons, in this paper, $p=1, \ldots , r$ which encompasses what are elsewhere referred to as the `soliton' representations together with the `anti-soliton' representations (e.g., see \cite{CZ09a}). The anti-soliton of a species $p$ soliton is a species $h-p$ soliton. In particular, in $a_2^{\! (2)}$, $p=1$ is `the soliton' and $p=2$ `the anti-soliton'.}  for $u$ and for $v$ revealed that a $B$ defect gives $v$ a delay factor of \cite{BCZ04}
\begin{align}
z_p = \frac{ie^{-\theta} + d\omega^{\stackrel{\frac{p}{2}}{}}}{ie^{-\theta} + d\omega^{\stackrel{-\frac{p}{2}}{}}} \label{del1}
\end{align}
while a $B^T$ defect gives a delay factor of \cite{CZ07}
\begin{align}
\tilde{z}_p = \frac{ie^{-\theta} - \tilde{d}\omega^{\stackrel{-\frac{p}{2}}{}}}{ie^{-\theta} - \tilde{d}\omega^{\stackrel{\frac{p}{2}}{}}} \label{del2}
\end{align}
where $\theta$ is the rapidity of the soliton and $\omega=e^{\stackrel{\frac{2\pi i}{r+1}}{}}$. The species $p$ soliton and the species $h-p$ soliton receive different delays, meaning that neither $B$ nor $B^T$ type I defects are compatible with the folding considered in this paper. \\
\\
The framework for defects was extended in \cite{CZ09b,CZ10b} to include integrable defects for $a_2^{\! (2)}$ by means of what is referred to as a type II defect with an ansatz of the form
\begin{align}
\mathcal{L} = \theta(-x)\mathcal{L}_u + \theta(x)\mathcal{L}_v + \delta(x)\left(- 2(u-v)\dot{\chi} - D(u,v,\chi) \right) \label{lag3}
\end{align}
where $u$ and $v$ are now scalar fields in either $a_1$ or $a_2^{\! (2)}$ (the form of $D$ depends upon which theory is being looked at) while $\chi$ is a scalar field which only exists at the defect, known as an auxiliary field. \\
\\
It was also noticed that a system containing a type I $B$ defect and a type I $B^T$ defect with the same parameter $d$ in $a_2^{\! (1)}$ would give the same overall time delay to the soliton and antisoliton\footnote{P.~Bowcock, private communication. Also mentioned in \cite{CZ09b}}; as such, solitons possessing the $a_2^{\! (2)}$ symmetry in the field $u$ would also possess it in the field $v$. This opens up the possibility that \eqref{lag3} for $a_2^{\! (2)}$ arises from fusing two type I $a_2^{\! (1)}$ defects (this was also noted in terms of B\"acklund transformations in \cite{CZ09b}). This possibility is formulated and generalised to all ATFTs obtainable by folding \ar in this paper.
\\
\\
Section 2 describes the foldings of \ar used in this paper (such foldings were also considered in \cite{KS}) and describes how soliton solutions of the folded theories may be obtained from \ar solitons using these folding processes. The relevant Hirota tau functions \cite{Hiro} are obtained and compared to the results of McGhee \cite{McGhee}. Section 3 describes the construction of a defect in \ar that is compatible with these folding processes. It is shown that the folded defect conserves momentum and that solitons in the folded theories retain their forms when passed through the defect. Section 4 gives the conclusions and outlook for this discourse.

\section{Folding and solitons}

In this section a folding process, previously considered in \cite{KS}, is formulated and used to construct soliton solutions for the folded theories $c_n^{(1)}$, $d_n^{(2)}$ and $a_{2n}^{\! (2)}$. This process for constructing solitons of $d_n^{(2)}$ and $a_{2n}^{\! (2)}$ using \ar does not appear elsewhere in the literature, though it may have been considered before\footnote{G.~M.~T.~Watts, private communication.}. Utilising \ar makes constructing multisolitons in the folded theories a much simpler task, this is demonstrated in section \ref{taus}.

\subsection{Folding \ar} \label{folding}

As this paper places its foundations in the firm ground of \ar, it is useful to note a few properties of \ar. The roots obey 
\begin{align}
\alpha_i \cdot \alpha_j = 2\delta_{ij} - \delta_{i(j+1)} - \delta_{i(j-1)} \label{cart}
\end{align}
for $i,j=0,\ldots,r$. It is useful to extend this range for $i$ and $j$ by identifying the indices modulo the Coxeter number $h$. For \ar, $h=r+1$, so e.g., $\alpha_{-2}\equiv \alpha_{r-1}$.  \\
\\
Conventionally, the fundamental highest weights $\{\lambda_i\}$ are given by
\begin{align*}
\lambda_i \cdot \alpha_j = \delta_{ij} \, ,\qquad i,j=1,\ldots,r
\end{align*}
with $\lambda_0 = 0$ so they form a dual basis to $\{\alpha_1,\ldots,\alpha_r\}$. Note then the relation
\begin{align*}
\alpha_i = 2\lambda_i - \lambda_{i-1} - \lambda_{i+1} \, .
\end{align*}
\\
The affine Toda field, $u$, lives on the \ar root space so a natural basis is given by the simple roots $\{\alpha_1,\ldots,\alpha_r\}$: $u=u_1 \alpha_1 + \ldots u_r \alpha_r$. Note then that $u_i = \lambda_i \cdot u$.
\\
\\
The three folded theories $c_n^{(1)}$, $d_n^{(2)}$ and $a_{2n}^{\! (2)}$ are all obtained here by identifying the roots of \ar pairwise, elucidating the work of Khastgir and Sasaki \cite{KS}. The roots of the folded theories are labelled by $\alpha'$; unfolded fields and denoted by $u$ and folded fields by $\phi$. 

\begin{itemize}

\item $a_{2n-1}^{\! (1)} \to c_n^{(1)}$: This folding is canonical (appearing in \cite{OT1}) and illustrated by figure \ref{atoc} with the black nodes representing roots of unit length and the extra root $\alpha_0$ placed on the left of the $a_{2n-1}^{\! (1)}$ part of the diagram. The root space identification made is
\begin{align*}
\alpha_i' = \frac{\alpha_i + \alpha_{h-i}}{2}
\end{align*}
where $h=2n$ and the there are two self identified roots, $\alpha_0' = \alpha_0$ and $\alpha_n' = \alpha_n$. \\
\\
The $a_{2n-1}^{\! (1)}$ affine Toda field u is folded to the $c_n^{(1)}$ affine Toda field $\phi$ by setting
\begin{align*}
u_i = u_{h-i} &= \frac{\phi_i}{2} \quad \text{for } i=1,\ldots,n-1 \\
u_n &= \phi_n \, .
\end{align*}
It is easily seen that the correct Lagrangian with the correct potential is obtained from these identifications.

\item $a_{2n-1}^{\! (1)} \to d_n^{(2)}$: This folding is non-canonical in the sense of \cite{KS}. It is illustrated by figure \ref{atod} where the black-in-white nodes represent roots of length $\frac{1}{\sqrt{2}}$ \footnote{The conventional normalisation for $d_n^{(2)}$ can be achieved by rescaling the roots $\alpha_i' \to \sqrt{2} \alpha_i'$. The affine Toda potential obtained from this folding is also non-standard, being twice the conventional potential - this factor of two can be removed in the action by an isotropic space-time rescaling: $x \to \frac{x}{\sqrt{2}}$, $t \to \frac{t}{\sqrt{2}}$.}. The root space identification required is
\begin{align*}
\alpha_i' = \frac{\alpha_i + \alpha_{h+1-i}}{2}
\end{align*}
where again $h=2n$ and $\alpha_0$ is chosen to be the lower-left root in figure \ref{atod}.
\\
\\
The $a_{2n-1}^{\! (1)}$ affine Toda field is folded to the $d_n^{(2)}$ affine Toda field by setting
\begin{align*}
u_{i+1} = u_{h-i} &= \frac{\phi_i}{2} \quad \text{for }i=1,\ldots,n-1 \\
u_{1} &= 0 \, .
\end{align*}

\item $a_{2n}^{\! (1)} \to a_{2n}^{\! (2)}$: This case is illustrated by figure \ref{atoa}, with the root space identification
\begin{align*}
\alpha_i' = \frac{\alpha_i + \alpha_{h-i}}{2}
\end{align*}
where the Coxeter number is now $h=2n+1$.\\
\\
The $a_{2n}^{\! (1)} \to a_{2n}^{\! (2)}$ folding is now achieved by setting
\begin{align*}
u_i = u_{h-i} = \frac{\phi_i}{2} \quad \text{for }i=1,\ldots,n  \, .
\end{align*}

\end{itemize}
\noindent
While \cite{KS} notes that there are other identifications of the roots than the choices above\footnote{Identifying $\alpha_i$ and $\alpha_{h+k-i}$ with general $k$ encompasses all of these possibilities. The choices made in this paper are $k=0$ for folding to $c_n^{(1)}$ and $a_{2n}^{\! (2)}$; while $k=1$ for folding to $d_n^{(2)}$.}, the resulting folded ATFTs are the same and the defect potentials that may be obtained are equivalent to those obtained in this paper. \\
\\
In what proceeds, the folding will be assumed to be achieved by setting $u_i = u_{h-i}$, fitting the pattern of folding to $c_n^{(1)}$ and $a_{2n}^{\! (2)}$. If folding to $d_n^{(2)}$ is instead desired then $h-i$ should be replaced by $h+1-i$ in what follows; the $d_3^{(2)}$ defect is explicitly given in section \ref{cases} to illustrate this point.

\begin{figure}
\centering
\begin{picture}(360,70)
 
\put(17,21){\circle{6}}
\put(40,3){\circle{6}}
\put(66,3){\circle{6}}
\put(92,3){\circle{6}}
\put(118,3){\circle{6}}
\put(144,3){\circle{6}}
\put(167,21){\circle{6}}
\put(40,39){\circle{6}}
\put(66,39){\circle{6}}
\put(92,39){\circle{6}}
\put(118,39){\circle{6}}
\put(144,39){\circle{6}}

\put(43,3){\line(1,0){20}}
\put(69,3){\line(1,0){20}}
\put(95,3){\line(1,0){20}}
\put(121,3){\line(1,0){20}}

\put(147,3){\line(4,3){20}}
\put(147,39){\line(4,-3){20}}

\put(17,24){\line(4,3){20}}
\put(17,18){\line(4,-3){20}}

\put(43,39){\line(1,0){20}}
\put(69,39){\line(1,0){20}}
\put(95,39){\line(1,0){20}}
\put(121,39){\line(1,0){20}}

\put(0,45){\small{$a_{2n-1}^{\! (1)}$}}

\put(175,21){\vector(1,0){20}}
\put(92,10){\vector(0,1){7}}
\put(92,32){\vector(0,-1){7}}

\put(204,21){\circle{6}}
\put(230,21){\circle*{6}}
\put(256,21){\circle*{6}}
\put(282,21){\circle*{6}}
\put(308,21){\circle*{6}}
\put(334,21){\circle*{6}}
\put(360,21){\circle{6}}

\put(207,22){\line(1,0){21}}
\put(207,20){\line(1,0){21}}

\put(233,21){\line(1,0){20}}
\put(259,21){\line(1,0){20}}
\put(285,21){\line(1,0){20}}
\put(311,21){\line(1,0){20}}

\put(357,22){\line(-1,0){21}}
\put(357,20){\line(-1,0){21}}

\put(200,38){\small{$c_{n}^{(1)}$}}

\end{picture}
\caption{$a_{2n-1}^{\! (1)} \to c_{n}^{(1)}$.}
\label{atoc}
\end{figure}
\noindent

\begin{figure}
 \centering
\begin{picture}(360,70)
\put(40,3){\circle{6}}
\put(66,3){\circle{6}}
\put(92,3){\circle{6}}
\put(118,3){\circle{6}}
\put(144,3){\circle{6}}
\put(170,3){\circle{6}}

\put(40,39){\circle{6}}
\put(66,39){\circle{6}}
\put(92,39){\circle{6}}
\put(118,39){\circle{6}}
\put(144,39){\circle{6}}
\put(170,39){\circle{6}}


\put(43,3){\line(1,0){20}}
\put(69,3){\line(1,0){20}}
\put(95,3){\line(1,0){20}}
\put(121,3){\line(1,0){20}}

\put(147,3){\line(1,0){20}}
\put(147,39){\line(1,0){20}}

\put(43,39){\line(1,0){20}}
\put(69,39){\line(1,0){20}}
\put(95,39){\line(1,0){20}}
\put(121,39){\line(1,0){20}}

\put(170,6){\line(0,1){30}}
\put(40,6){\line(0,1){30}}

\put(0,45){\small{$a_{2n-1}^{\! (1)}$}}

\put(175,21){\vector(1,0){20}}
\put(92,10){\vector(0,1){7}}
\put(92,32){\vector(0,-1){7}}

\put(204,21){\circle{6}}
\put(204,21){\circle*{3}}
\put(230,21){\circle*{6}}
\put(256,21){\circle*{6}}
\put(282,21){\circle*{6}}
\put(308,21){\circle*{6}}
\put(334,21){\circle*{3}}
\put(334,21){\circle{6}}

\put(207,22){\line(1,0){21}}
\put(207,20){\line(1,0){21}}

\put(233,21){\line(1,0){20}}
\put(259,21){\line(1,0){20}}
\put(285,21){\line(1,0){20}}

\put(331,22){\line(-1,0){21}}
\put(331,20){\line(-1,0){21}}

\put(200,38){\small{$d_n^{(2)}$}}

\end{picture}
\caption{$a_{2n-1}^{\! (1)} \to d_n^{(2)}$.}
\label{atod}
\end{figure}
\noindent

\begin{figure}
\centering
\begin{picture}(360,70)

\put(17,21){\circle{6}}
\put(40,3){\circle{6}}
\put(66,3){\circle{6}}
\put(92,3){\circle{6}}
\put(118,3){\circle{6}}
\put(144,3){\circle{6}}
\put(170,3){\circle{6}}

\put(40,39){\circle{6}}
\put(66,39){\circle{6}}
\put(92,39){\circle{6}}
\put(118,39){\circle{6}}
\put(144,39){\circle{6}}
\put(170,39){\circle{6}}

\put(43,3){\line(1,0){20}}
\put(69,3){\line(1,0){20}}
\put(95,3){\line(1,0){20}}
\put(121,3){\line(1,0){20}}

\put(147,3){\line(1,0){20}}
\put(147,39){\line(1,0){20}}

\put(17,24){\line(4,3){20}}
\put(17,18){\line(4,-3){20}}

\put(43,39){\line(1,0){20}}
\put(69,39){\line(1,0){20}}
\put(95,39){\line(1,0){20}}
\put(121,39){\line(1,0){20}}
\put(170,6){\line(0,1){30}}

\put(0,45){\small{$a_{2n}^{\! (1)}$}}

\put(175,21){\vector(1,0){20}}
\put(92,10){\vector(0,1){7}}
\put(92,32){\vector(0,-1){7}}

\put(204,21){\circle{6}}
\put(230,21){\circle*{6}}
\put(256,21){\circle*{6}}
\put(282,21){\circle*{6}}
\put(308,21){\circle*{6}}
\put(334,21){\circle*{6}}
\put(360,21){\circle*{3}}
\put(360,21){\circle{6}}

\put(207,22){\line(1,0){21}}
\put(207,20){\line(1,0){21}}

\put(233,21){\line(1,0){20}}
\put(259,21){\line(1,0){20}}
\put(285,21){\line(1,0){20}}
\put(311,21){\line(1,0){20}}

\put(357,22){\line(-1,0){21}}
\put(357,20){\line(-1,0){21}}

\put(200,38){\small{$a_{2n}^{\!(2)}$}}

\end{picture}
\caption{$a_{2n}^{\! (1)} \to a_{2n}^{\! (2)}$.}
\label{atoa}
\end{figure}
\noindent

\subsection{\ar solitons and folding}

This section details the construction of solitons in the folded theories, all of which are \ar solitons possessing a certain symmetry. In this context, \ar solitons must be introduced first.
\\
\\
Hirota methods \cite{Hiro} are used to find soliton solutions in the ATFTs. For this paper the starting point will be \ar, as it is the theories obtained from folding this that are of interest. For a general ATFT the soliton solutions may be written in the form \cite{MM}
\begin{align}
u = - \frac{1}{\beta}\sum_{j=0}^r \eta_j \alpha_j \ln \tau_j  \label{solgen}
\end{align}
where $\eta_i=\frac{2}{\alpha_i \cdot \alpha_i}$ with no sum implied and the ATFT corresponds to an affine Dynkin diagram with $r+1$ nodes. \\
\\
The tau functions $\{\tau_j \}$ depend only upon the root data (i.e., which ATFT is being considered) and do not depend upon the coupling. In order to find the tau functions \eqref{solgen} is used in the equation of motion along with a decoupling \cite{Holl}. In general the equation to be solved is
\begin{align}
 \eta_i \left( \ddot{\tau}_i \tau_i - \dot{\tau}_i^2 - \tau_i'' \tau_i + \tau_i'^2 \right) - n_i \left( \prod_{j=0}^r \tau_j^{-\eta_j \alpha_j \cdot \alpha_i} - 1 \right) \tau_i^2 = 0 \; . \label{soleom}
\end{align}
\\
For \ar, $\eta_i=n_i=1$ for all $i$, so \eqref{solgen} becomes
\begin{align}
u = - \frac{1}{\beta} \sum_{j=0}^r \alpha_j \ln \tau_j  \label{sola}
\end{align}
while the equation the tau functions must obey simplifies greatly to
\begin{align}
\ddot{\tau}_i \tau_i - \dot{\tau}_i^2 - \tau_i'' \tau_i + \tau_i'^2  = \tau_{i-1}\tau_{i+1} - \tau_i^2 \; . \label{aeom}
\end{align}
\\
\! \! \!In the cases of interest folding is achieved by setting $u_i = u_{h-i}$ which is tantamount to having $\tau_i = \tau_{h-i}$ (N.B. it is to be understood that $h-i$ should be replaced by $h+1-i$ if the case being considered is $d_n^{(2)}$). Any \ar soliton with this property is also a soliton of the folded theory. The single soliton solutions for $d_n^{(2)}$ and $a_{2n}^{\! (2)}$ already known from \cite{McGhee} will be recovered from this different approach (The construction of the $c_n^{(1)}$ single solitons in \cite{MM} used the same approach as this).
\begin{table}
\centering
\begin{align*}
\begin{array}{|c|c|c|c|}
\hline
 & c_n^{(1)} & d_n^{(2)} & a_{2n}^{\! (2)} \\
 \hline
 \tau_0' & \tau_{0} & (\tau_1)^{\frac{1}{2}} & \tau_{0} \\
 \hline
 \tau_i' \; \; i\neq 0,n-1,n & \tau_{i} & \tau_{1+i} & \tau_{i} \\
 \hline
 \tau_{n-1}' & \tau_{n-1} & (\tau_{n})^{\frac{1}{2}} & \tau_{n-1} \\
 \hline
 \tau_n' & \tau_{n} & - & (\tau_{n})^{\frac{1}{2}} \\
 \hline 
 \end{array}
 \end{align*}
\caption{Tau function identifications in the folded theories.}
\label{foldtable}
\end{table}
\noindent

\subsection{Tau functions} \label{taus}

A formula is known for the tau functions of the $N$ soliton solution in \ar which is \cite{Holl}
\begin{align}
 \tau_j = \sum_{\mu_1 = 0}^1 \ldots \sum_{\mu_N = 0}^1 \exp \left[ \sum_{i=1}^N \mu_i \ln \left( \omega^{p_i j}E_{p_i} \right) + \sum_{1\leq i < j \leq N} \mu_i \mu_j \ln 
\left( A^{(p_i p_j)} \right) \right] \; . \label{multisol}
\end{align}
\\
In \eqref{multisol}, $p_i$ denotes the species of soliton $i$ ($p_i = 1, \ldots ,r$) and $\omega = e^{\frac{2 \pi i}{h}}$ with $h=r+1$ the Coxeter number. The spacetime dependence of the tau function is found in $E_{p_i}$:
\begin{align}
E_{p_i} = e^{a_{p_i} x - b_{p_i} t + c_{p_i}} && \text{with} \quad a_{p_i} = m_{p_i} \cosh \theta_i \quad, \quad b_{p_i} = m_{p_i} \sinh \theta_i \; . \label{eppy}
\end{align}
\\
In \eqref{eppy} the rapidity of soliton $i$ is denoted by $\theta_i$ while $c_{p_i}$ is a complex parameter - the imaginary part of which determines the topological charge of the soliton \cite{McGh} and which must be chosen to avoid singularities in the soliton solution \eqref{sola}. The quantity $m_{p_i} = 2 \sin \left( \frac{\pi {p_i}}{h} \right)$ is the positive square root of the $p$th eigenvalue of the Cartan matrix and the mass of soliton $i$ is then $M_{p_i}^2 = \frac{2 h}{\beta^2}m_{p_i}^2$ \cite{Holl,MM}. \\
\\
The other quantities in the tau functions \eqref{multisol} are the interaction parameters $\{ A^{(p_i p_j)} \}$, they are given by 
\begin{align}
 A^{(pq)} = - \frac{(a_p - a_q)^2 - (b_p - b_q)^2 - m_{p-q}^2}{(a_p + a_q)^2 - (b_p + b_q)^2 - m_{p+q}^2} \; . \label{intparam} 
\end{align}
\\
To find the tau functions for single solitons of species $p$ in the folded theories take a two soliton \ar solution ($N=2$) with $p_1 = p$ and $p_2 = h-p$ in \eqref{multisol}. The two \ar solitons must then be given the same centre of mass, $\mathcal{R}(c_{p_1}) = \mathcal{R}(c_{p_2})$, as well as the same rapidity, $\theta_1 = \theta_2$. This means that $E_{p_1}$ and $E_{p_2}$ may differ only in $\mathcal{I}(c)$. \\
\\
With $\theta_1 = \theta_2$, \eqref{intparam} becomes
\begin{align}
A^{(p(h-p))} =\cos^2 \left(\frac{\pi p}{r+1}\right) \equiv A \; . \label{intparam2}
\end{align}
\\
Thus, the tau functions compatible with folding (to a one soliton folded solution) possess the form\footnote{For the case of $d_n^{(2)}$ this becomes $\tau_j = 1 + \left( \omega^{pj} + \omega^{p(h+1-j)} \right) E_p + A \omega^{p}E_p^2$.}
\begin{align}
\tau_j = 1 + \left( \omega^{pj} + \omega^{p(h-j)} \right) E_p + A E_p^2  \; . \label{foldsol}
\end{align}
\\
It can be shown that these folded solitons are the same as those found in \cite{McGhee} with the identifications in table \ref{foldtable}. 
\\
\\
It is only for \ar that a formula like \eqref{multisol} is known. Once the basic tau functions for one soliton solutions in the folded theories are known, multisoliton solutions in these models may be constructed directly; however, this requires knowledge of the generally complicated interaction parameters of the folded model. This problem may be obviated by instead constructing these multisolitons in the \ar model. This was noted in \cite{ZC}, but was only applied to $c_n^{(1)}$. In particular, the two soliton solution in the folded theory takes the form
\begin{align}
\tau_j &= 1 + \left(\omega^{pj} + \omega^{p(h-j)}\right) E_p + \left(\omega^{qj} + \omega^{q(h-j)}\right)E_q + A^{(12)}E_p^2 + A^{(34)}E_q^2 \nonumber \\
& \qquad + A^{(13)} \left(\omega^{pj + qj} + \omega^{p(h-j) + q(h-j)} \right)E_p E_q + A^{(14)} \left(\omega^{pj + q(h-j)} + \omega^{p(h-j) + qj} \right) E_p E_q \nonumber \\
& \qquad + A^{(12)}A^{(13)}A^{(14)} \left( \omega^{qj} + \omega^{q(h-j)}\right) E_p^2 E_q \nonumber \\
& \qquad + A^{(34)}A^{(13)}A^{(14)}\left( \omega^{pj} + \omega^{p(h-j)}\right) E_p E_q^2 \nonumber \\
& \qquad + A^{(12)}A^{(34)}\left(A^{(13)}\right)^2 \left(A^{(14)}\right)^2 E_p^2 E_q^2 \; . \label{twosol}
\end{align}
\\
Note that \eqref{twosol} contains four interaction parameters - a fact that is not obvious, should one wish to construct folded solitons using the folded theory as a starting point.
\\
\\
Using that $a_p = m_p \cosh \theta_1, a_q = m_q \cosh \theta_2, b_p = m_p \sinh \theta_1, b_q = m_q \sinh \theta_2$ and denoting the two rapidities by $\theta_1 = \theta + \psi$ and $\theta_2 = \theta - \psi$ gives, in particular, the interaction parameter
\begin{align}
 A^{(13)} &= \frac{m_{p-q}^2 + (m_p + m_q)^2 \sinh^2 \psi - (m_p - m_q)^2 \cosh^2 \psi}{(m_p + m_q)^2 \cosh^2 \psi - (m_p - m_q)^2 \sinh^2 \psi - m_{p+q}^2}  \; . \label{aonethree} 
\end{align}
\\
Among the two soliton solutions there are two interesting cases that can occur when the the relative rapidity $\theta_1 - \theta_2 = 2 \psi$ between the solitons is imaginary:
\begin{itemize}

\item The solitons possess fusing rules, which are just $a_r^{\! (1)}$ fusing rules. Fusion of the solitons occurs when the denominator of $A^{(13)}$ in equation \eqref{aonethree} vanishes (one should first make the redefinitions $E_p \to \left(A^{(13)}\right)^{-\frac{1}{2}}E_p$ and $E_q \to \left(A^{(13)}\right)^{-\frac{1}{2}}E_q$ in equation \eqref{twosol}). This occurs when $\psi = \pm i \frac{\pi (p+q)}{2(r+1)} \equiv \pm i \tilde{\psi}$. The resulting tau functions describe a species $s=p+q$ single folded soliton with rapidity $\tilde{\theta} = \theta + i\frac{\pi (p-q)}{2(r+1)}$.

\item The existence of breather solutions in ATFT has been known for some time \cite{OTU2} and solutions have been considered in Hirota form for \ar \cite{HIM} and $d_4^{(1)}$ \cite{Iskandar}. For equation \eqref{twosol} to describe a folded breather the constituent solitons must be of the same species, $p=q$, with the same centre of mass $\mathcal{R}(c_1) = \mathcal{R}(c_2)$ (where $E_p \to E_1$ and $E_q \to E_2$) and with imaginary rapidity difference, $\psi = i \tilde{\psi}$ - $2\tilde{\psi}$ must be less than the fusing angle. Note that in the cases of $d_n^{(2)}$ and $a_{2n}^{\! (2)}$ these breather tau functions are also the tau functions of particular breathers in $d_s^{(1)}$.

\end{itemize}

\section{Folding defects}

In this section defects are constructed for folded \ar systems. The type I defects described in \cite{BCZ04} do not possess the symmetry required for folding but two kinds exist (referred to as $B$ and $B^T$ defects in the introduction section) which are conjugate in that bringing them together, `fusing' them, results in precisely the kind of defect that can be folded, provided that the defect rapidities are appropriately related. Two methods will be used to test the possible integrability of the folded defect system. Firstly, conservation of a modified energy and momentum will be shown, provided that the defect rapidities are identified in a certain way. Na\"ively the expectation is that momentum should not be conserved in the presence of a defect due to the breaking of spatial translation invariance, however it was shown in \cite{CZ09a} for type I \ar defects that when momentum is conserved, it implies all of the conditions for classical integrability found by the Lax pair approach in \cite{BCZ04}. Secondly, it is shown that the solitons of the folded theory preserve their forms when passed through the folded defect. The combination of momentum conservation and solitons preserving their forms strongly suggest the existence of classically integrable $c_n^{(1)}$, $d_n^{(2)}$ and $a_{2n}^{\! (2)}$ defects. This is the main point of this paper.
\\
\\
Consider a system with a type I $B$ defect at $x=0$ and a type I $B^T$ defect at $x=a>0$. The Lagrange density describing this system may be written as
\begin{align}
\mathcal{L} &= \theta(-x) \mathcal{L}_u + \delta(x)\left(\frac{1}{2}uA\dot{u} + uB\dot{\chi} + \frac{1}{2}\chi A\dot{\chi} - 
D^{(1)}(u,\chi)\right) + \theta(x)\theta(a-x) \mathcal{L}_{\chi} \nonumber \\
 & \quad  + \delta(x-a) \left(-\frac{1}{2}\chi A\dot{\chi} + \chi B^T \dot{v} - \frac{1}{2}v A\dot{v} - D^{(2)}(\chi,v)\right) + \theta(x-a)\mathcal{L}_v \label{unfusedlag}
\end{align}
where $A=1-B$ and $B+B^T=2$ with $\mathcal{L}_u$ being a Lagrange density in the form of \eqref{lag1}; $\mathcal{L}_v$ and $\mathcal{L}_{\chi}$ similarly. The defect potentials are then given by
\begin{align}
D^{(1)} &= D^{(1)-} + D^{(1)+} = \sum_{j=0}^r f_j + \sum_{j=0}^r g_j \label{done} \\
D^{(2)} &= D^{(2)-} + D^{(2)+} = \sum_{j=0}^r \tilde{f}_j + \sum_{j=0}^r \tilde{g}_j \label{dtwo}
\end{align} 
where the terms with $f_i$ and those with $\tilde{f}_i$ possess negative helicity and those with $g_i$ or $\tilde{g}_i$ are of positive helicity:
\begin{align}
f_i &= d e^{\stackrel{{\frac{1}{2}\alpha_i \left(B^T u + B\chi \right)}}{}} \; , \; g_i = d^{-1} e^{\stackrel{{\frac{1}{2}\alpha_i B \left(u - \chi \right)}}{}} \; , \; \tilde{f}_i = \tilde{d} e^{\stackrel{{\frac{1}{2}\alpha_i \left(B^T v + B\chi \right)}}{}} \; , \;
\tilde{g}_i = \tilde{d}^{-1} e^{\stackrel{{\frac{1}{2}\alpha_i B \left(v - \chi \right)}}{}} \; . \label{notation} 
\end{align}
\\
The defects may be brought together, or fused, at the Lagrangian level by taking $a \to 0$ in \eqref{unfusedlag} (c.f., the sine-Gordon case in \cite{CZ09b,CZ10a}) resulting in
\begin{align}
 \mathcal{L} &= \theta(-x)\mathcal{L}_u +  \theta(x)\mathcal{L}_v \nonumber \\
 &\quad + \delta(x) \left( \frac{1}{2}uA\dot{u} + uB\dot{\chi}- vB\dot{\chi} -  \frac{1}{2}vA\dot{v} - D^{(1)}- D^{(2)} \right) \; . \label{alag}
 \end{align}
 \\
There no longer exists any bulk for the field $\chi$; it is effectively trapped in the defect and hence may be referred to as an `auxiliary field'. Note that in order for \eqref{alag} to match the type II ansatz of \cite{CZ10b} the auxiliary field must be redefined to account for the presence of self-coupling defect terms for the bulk fields $u$ and $v$. 
\\
\\
The Euler--Lagrange equations of \eqref{alag} give now the bulk equations for $u$ and $v$ as well as the defect conditions at $x=0$
 \begin{align}
 u' &= A\dot{u} + B\dot{\chi} - D_u  \label{audash} \\
 v' &= A\dot{v} + B\dot{\chi} + D_v  \label{avdash} \\
 B^T \dot{u} &+ D^{(1)}_{\chi}= B^T \dot{v} - D^{(2)}_{\chi} \label{audot} 
 \end{align}
where $D$ without a superscript refers to the fused defect potential $D=D^{(1)}+D^{(2)}$. The fused defect thus gives three vector equations as the defect conditions while the unfused system has four; however, the delay factors received by solitons passing through the defect are unchanged by the fusing process.
\\
\\
A modified energy is conserved by the fused defect in a simple way. The conserved energy is just the Hamiltonian of the fused defect system and is given by $E+D$ where 
\begin{align*}
E = \int_{-\infty}^{0} \frac{1}{2}\dot{u}\cdot \dot{u} + \frac{1}{2}u' \cdot u' + U \; \mathrm{d}x + \int_{0}^{\infty} \frac{1}{2}\dot{v}\cdot \dot{v} + \frac{1}{2}v' \cdot v' + V \; \mathrm{d}x
\end{align*}
is the bulk contribution to the energy and $D$ is the defect contribution. \\
\\
It is not so obvious from the outset that there should be a conserved momentum in the case of the fused defect. The bulk contribution to the momentum is given by the integral of $T^{01}$ of the stress tensor derived from \eqref{alag}, so
\begin{align*}
P = \int_{-\infty}^0 \dot{u}\cdot u' \; \mathrm{d}x + \int_0^{\infty} \dot{v}\cdot v' \; \mathrm{d}x \; .
\end{align*}
\\
Taking the time derivative and using the bulk equations of motion gives
\begin{align*}
\dot{P} = \frac{1}{2}\dot{u}\cdot \dot{u} + \frac{1}{2}u' \cdot u' - U - \frac{1}{2}\dot{v}\cdot \dot{v} - \frac{1}{2}v' \cdot v' + V \; |_{x=0}  
\end{align*}
so long as the fields and potentials are constant (in vacuum) at spatial infinity. Using the defect conditions \eqref{audash}, \eqref{avdash} and \eqref{audot} one can then show that
\begin{align}
\dot{P} = - \left( \dot{D}^- - \dot{D}^+ \right)  \label{momresult}
\end{align}
and thus that $P + D^- - D^+$ is a conserved quantity.
\\
\\
Note that an analogous analysis holds if in the first instance in \eqref{unfusedlag} the $B$ defect is taken to be to the right of the $B^T$ defect. This perhaps should come as no surprise as one may appeal to commutability given that the defect conditions describe B\"acklund transformations \cite{BCZ04}.

\subsection{The folded defect}

By taking a type I \ar defect fused to the defect that is its image under folding, a defect, \eqref{alag}, has been constructed which is symmetric under folding. For this to describe a defect in the folded theory  the bulk fields $u$ and $v$ must be folded, but at the defect it is not clear what should be done with the auxiliary field $\chi$. Consideration of the time delays of soliton solutions suggests that the field $\chi$ should not be folded - this was seen in the case of $a_2^{\! (1)} \to a_2^{\! (2)}$ in \cite{CZ09b}.
\\
\\
One simplification that occurs in folding the Lagrangian \eqref{alag} is that the self-coupling kinetic terms at the defect vanish because $\alpha_i' B \alpha_j' = \alpha_i' \cdot \alpha_j'$. Hence, since $A=1-B$, 
\begin{align}
\phi A \dot{\phi} = \psi A \dot{\psi} = 0   \label{noselfcoup}
\end{align} 
where the folding is denoted by $u \to \phi$ and $v \to \psi$. This means that the Lagrangian \eqref{alag} folds to
\begin{align}
\mathcal{L} &= \theta (-x) \mathcal{L}_{\phi} + \theta (x) \mathcal{L}_{\psi} \nonumber \\
& \quad + \delta(x) \left( \phi B \dot{\chi} - \psi B \dot{\chi} - D^{(1)} (\phi , \chi) - D^{(2)}(\chi, \psi) \right)  
\label{foldlag}
\end{align}
which fits the type II framework of \cite{CZ09b} without requiring redefinition of $\chi$.
\\
\\
In vector form the Euler--Lagrange equations are thus
\begin{align}
\phi' &= {}_p B \dot{\chi} - D_{\phi}  \label{phidash} \\
\psi' &= {}_p B \dot{\chi} + D_{\psi}  \label{psidash} \\
B^T \dot{\phi} &+ D^{(1)}_{\chi} = B^T \dot{\psi} - D^{(2)}_{\chi} \; . \label{phidot}
\end{align}
\\
The subscript $p$ found in \eqref{phidash} and \eqref{psidash} in front of $B \dot{\chi}$ denotes `projected' and indicates that equations \eqref{phidash} and \eqref{psidash} only make sense when projected onto the folded root space.
\\
\\
Examination of the components of \eqref{phidot} gives the algebraic constraints
\begin{align*}
D_{\chi_i} + D_{\chi_{h-1-i}} = 0
\end{align*}
which, using $D^-_{\chi} = \sum_j \frac{1}{2}B^T \alpha_j \left(f_j + \tilde{f}_j \right)$ and $D^+_{\chi} = -\sum_j \frac{1}{2}B^T \alpha_j \left(g_j + \tilde{g}_j \right)$ may be put in the form
\begin{align}
f_i - f_{h-i} + \tilde{f}_i - \tilde{f}_{h-i} - g_i + g_{h-i} - \tilde{g}_i + \tilde{g}_{h-i} = 0 \; . \label{algconab}
\end{align}
\\
The algebraic constraints \eqref{algconab} relate terms of different helicities and so would appear to be incompatible with momentum conservation (since the momentum is a difference between positive and negative helicity terms). The solution is to relate the two defect parameters $d$ and $\tilde{d}$ in such a way as to prevent \eqref{algconab} from mixing helicities, resulting in the identification
\begin{align}
\tilde{d} = \pm d \, . \label{dident}
\end{align}
\\
With the above identification helicity preserving algebraic constraints are found, i.e., 
\begin{align}
D^-_{\chi_i} + D^-_{\chi_{h-1-i}} &= 0 \\
D^+_{\chi_i} + D^+_{\chi_{h-1-i}} &= 0
\end{align}
resulting in\footnote{Recall that $h$ should be replaced by $h+1$ in these conditions when folding to $d_n^{(2)}$.}
\begin{align}
f_i - f_{h-i} + \tilde{f}_i - \tilde{f}_{h-i} &= 0 \label{algcona} \\
g_i - g_{h-i} + \tilde{g}_i - \tilde{g}_{h-i} &= 0 \label{algconb}  \; .
\end{align}
\\
It is easily shown that $E+D$, where $E$ is the folded bulk energy, continues to be a conserved energy for the folded defect theory. It is shown here that, after folding, the quantity $P + D^- - D^+$ also continues to be conserved - where $P$ and $D$ involve now the folded fields $\phi$ and $\psi$ rather than $u$ and $v$.
\\
\\
The momentum conservation argument by itself is the most involved of the calculations presented in the paper and relies on the helicity conserving algebraic constraints \eqref{algcona} and \eqref{algconb}. The presence of the folded defect modifies the bulk momentum such that
\begin{align}
\dot{P} = \frac{1}{2}\left( \phi' \cdot \phi' + \dot{\phi}\cdot \dot{\phi} - \psi' \cdot \psi' - \dot{\psi}\cdot \dot{\psi}\right) -\Phi + \Psi  \label{pdot}
\end{align}
where $\Phi$ and $\Psi$ are the folded bulk potentials. A modified momentum $P+C$ is conserved if the right-hand side of \eqref{pdot} can be written as $-\frac{\mathrm{d} C}{\mathrm{d}t}$, so the aim is to show that this is the case and that $C = D^- - D^+$. The first and third terms on the right-hand side of \eqref{pdot} can be re-expressed, by taking the difference of the squares of \eqref{phidash} and \eqref{psidash}, as
\begin{align}
\phi' \cdot \phi'-\psi' \cdot \psi' = - 2\dot{\chi}\left(B^T D_{\phi}+B^T D_{\psi}\right) + D_{\phi}^2 - D_{\psi}^2 \; . \label{dashed}
\end{align}
\\
At this stage progress can be made by anticipating the form of the final answer to be $\dot{P} = - \left( \dot{D}^- - \dot{D}^+ \right)$, which requires the term $-\dot{\chi} \left( D_{\chi}^- - D_{\chi}^+ \right)$. The only place that $\dot{\chi}$ may appear in \eqref{pdot} stems from $\phi' \cdot \phi'-\psi' \cdot \psi'$ and so the conclusion is that
\begin{align}
B^T D^-_{\phi} + B^T D^-_{\psi} &= D^-_{\chi} \label{dchiminus} \\
B^T D^+_{\phi} + B^T D^+_{\psi} &= -D^+_{\chi} \; . \label{dchiplus}
\end{align}
\\
By using the relations
\begin{align}
 D^-_{\phi} = \sum_{j=0}^{h-1} \frac{1}{4}\left( B\alpha_j + B^T \alpha_{h-j} \right) f_j & & 
 D^+_{\phi} = \sum_{j=0}^{h-1} \frac{1}{4}\left( B^T\alpha_j + B \alpha_{h-j} \right) g_j  \label{dphi} \\
 D^-_{\psi} = \sum_{j=0}^{h-1} \frac{1}{4}\left( B\alpha_j + B^T \alpha_{h-j} \right) \tilde{f}_j & & 
 D^+_{\psi} = \sum_{j=0}^{h-1} \frac{1}{4}\left( B^T\alpha_j + B \alpha_{h-j} \right) \tilde{g}_j  \label{dpsi}
 \end{align}
and the constraints \eqref{algcona}, \eqref{algconb} along with the fact that $B + B^T = 2$, it can be shown that \eqref{dchiminus} and \eqref{dchiplus} are true. Since \eqref{dchiminus} and \eqref{dchiplus} are both true then \eqref{phidot} may be rewritten, noting that $B^T$ is invertible, as
\begin{align}
\dot{\phi} + D^-_{\phi} - D^+_{\phi} = \dot{\psi} - D^-_{\psi} + D^+_{\psi} \; . \label{dotphipsi}
\end{align}
\\
The equation \eqref{dotphipsi} may then be squared on both sides to give $\dot{\phi}\cdot \dot{\phi} - \dot{\psi}\cdot \dot{\psi}$ thus reducing \eqref{pdot} to
\begin{align*}
\dot{P} = -\dot{D}^- + \dot{D}^+ + 2D^-_{\phi}D^+_{\phi} - 2D^-_{\psi}D^+_{\psi} - \Phi + \Psi \; . 
\end{align*}
\\
This is almost what is sought, only requiring that
\begin{align}
2D^-_{\phi}D^+_{\phi} - 2D^-_{\psi}D^+_{\psi} = \Phi - \Psi \; . 
\end{align}
\\
Using \eqref{dphi} and \eqref{dpsi} along with $B+B^T = 2$ it is seen that
\begin{align*}
2 D_{\phi}^- D_{\phi}^+ - 2 D_{\psi}^- D_{\psi}^+ &= \frac{1}{2}\sum_{i,j} \left(f_i g_j - \tilde{f}_i \tilde{g}_j\right) \left(\alpha_i B^T \alpha_j \right) 
 + \sum_{i,j} M_{ij} \left(f_i g_j - \tilde{f}_i \tilde{g}_j\right) \\
&= \Phi - \Psi + \sum_{i,j} M_{ij} \left(a_i b_j - \tilde{a}_i \tilde{b}_j\right)
\end{align*}
where
\begin{align*}
M_{ij} = \frac{1}{8} \left( \alpha_i - \alpha_{h-i} \right) B B^T \left( \alpha_{h-j} - \alpha_j \right) \; .
\end{align*}
\\
It can be shown using the algebraic constraints \eqref{algcona} and \eqref{algconb} that this extra term vanishes and so the conserved momentum has the expected form, i.e., $P + D^- - D^+$, provided at least that the two defect parameters are related by $\tilde{d} = \pm d$. Momentum conservation is a strong constraint, perhaps even implying integrability, as is the fact that solitons can preserve their form in passing through the defect. It is now shown that both approaches result in the same relation between $\tilde{d}$ and $d$.
\\
\\
Consider passing solitons through the fused defect before folding. The soliton to the left of the defect is chosen to possess the symmetry of a single soliton of species $p$ of the folded theory, so\footnote{Or $\tau_j^u = 1 + \left( \omega^{pj} + \omega^{p(h+1-j)} \right) E_p + A \omega^{p} E_p^2$, should one wish to fold to $d_n^{(2)}$ instead of $c_n^{(1)}$ or $a_{2n}^{\! (2)}$.} 
\begin{align}
\tau_j^u = 1 + \left( \omega^{pj} + \omega^{p(h-j)} \right) E_p + A E_p^2 \label{u2}
\end{align}
where $A= \cos^2 \left( \frac{\pi p}{r+1} \right)$. Evolving this soliton through the defect using the defect equations \eqref{audash}, \eqref{avdash} and \eqref{audot} it picks up delay factors giving
\begin{align}
\tau_j^{\chi} &= 1 + \left(\omega^{pj} z_p + \omega^{p(h-j)} z_{h-p}\right) E_p + A z_p z_{h-p} E_p^2 \label{chi2} \\
\tau_j^v &= 1 + \left(\omega^{pj} z_p \tilde{z}_p  + \omega^{p(h-j)} z_{h-p} \tilde{z}_{h-p}\right) E_p + A z_p z_{h-p} \tilde{z}_p \tilde{z}_{h-p} E_p^2 \label{v2}
\end{align}
with the delay factors given by
\begin{align}
z_p = \frac{ie^{-\theta} + d\omega^{\stackrel{\frac{p}{2}}{}}}{ie^{-\theta} + d\omega^{\stackrel{-\frac{p}{2}}{}}} & & z_{h-p} &= \frac{ie^{-\theta} - d\omega^{\stackrel{-\frac{p}{2}}{}}}{ie^{-\theta} - d\omega^{\stackrel{\frac{p}{2}}{}}}   \nonumber  \\
\; \tilde{z}_p = \frac{ie^{-\theta} - \tilde{d}\omega^{\stackrel{-\frac{p}{2}}{}}}{ie^{-\theta} - \tilde{d}\omega^{\stackrel{\frac{p}{2}}{}}} & & \tilde{z}_{h-p} &= \frac{ie^{-\theta} + \tilde{d}\omega^{\stackrel{\frac{p}{2}}{}}}{ie^{-\theta} + \tilde{d}\omega^{\stackrel{-\frac{p}{2}}{}}} \; .   \label{delays}
\end{align}
\\
It is clear that $u$ may be folded as \eqref{u2} represents a folded soliton, this is the choice made initially. Equally clear is that in general $\tau_j^{\chi} \neq \tau_{h-j}^{\chi}$, since $z_p \neq z_{h-p}$ except in the single case that $p=n$ in $a_{2n-1}^{\! (1)}$, so it is not possible to fold the auxiliary field $\chi$. 
\\
\\
The aim of this soliton transmission argument is to find the circumstances under which the transmitted soliton, $v$, has the folding symmetry, so $\tau_j^v = \tau_{h-j}^v$. This is enough to show that the folded defect is able to preserve the forms of folded solitons - giving a very strong condition which may imply integrability. The reason for this claim is that any configuration that satisfies \eqref{audash}, \eqref{avdash} and \eqref{audot}, with the bulk fields $u$ and $v$ possessing the folding symmetry, also satisfies the folded defect conditions \eqref{phidash}, \eqref{psidash} and \eqref{phidot}. The requirement then is that $\tau_j^v = \tau_{h-j}^v$ and so it must be the case that
\begin{align*}
z_p \tilde{z}_p = z_{h-p} \tilde{z}_{h-p}
\end{align*}
in which case every $z$ may be absorbed into the definition of $E_p$ as a time delay and phase shift. Note that this condition is the same condition as having the \ar single soliton species $p$ and species $h-p$ solutions receiving the same delay factor through the fused defect. Thus, the condition for the soliton on the right of the defect to be in the folded theory is
\begin{align*}
0 = z_p \tilde{z}_p - z_{h-p} \tilde{z}_{h-p} = \frac{1}{\text{denom.}}\left[ e^{2 \theta} \left( \omega^p - \omega^{-p} \right) \left( d^2 - \tilde{d}^2 \right) \right]
\end{align*}
where `denom.' is the common denominator obtained by multiplying all of the denominators of \eqref{delays} together. So the defect represented by the Lagrangian \eqref{foldlag} is only likely to be integrable if $\tilde{d} = \pm d$, as this is what is required for the soliton solution \eqref{v2} to be compatible with folding. Therefore, there are two possibilities then that give folded solitons to the right of the defect
\begin{itemize}
\item When $\tilde{d} = - d$ it is the case that $z_p \tilde{z}_p = z_{h-p} \tilde{z}_{h-p} = 1$. i.e., all of the solitons receive a trivial time delay. Indeed in this case if $\psi = \phi$ is imposed then the defect part of the Lagrangian \eqref{foldlag} vanishes - so there is no defect there. The interpretation of this is that the second defect is the anti-defect of the first - fusing them causes annihilation.

\item When $\tilde{d} = d$ there is a delay factor (different for each $p$) so this should represent a \emph{bona fide} defect which does not destroy the form of the solitons (a strong constraint, suggesting that the defect is integrable). Note that if $\theta$ and $d$ are real then the delay factor is real too - there is a non-trivial time delay or advance and no change of topological charge. \\
\end{itemize}
\noindent
It is noteworthy that condition \eqref{dident} links the soliton time delay argument to the momentum conservation one. The condition ensures that solitons retain their form in the folded defect model and also ensures that the algebraic constraints do not mix helicities - something which the momentum conservation argument relies upon. 

\subsection{Specific cases} \label{cases}

The Lagrangian \eqref{foldlag} represents defects for the folded theories $c_n^{(1)}$, $d_n^{(2)}$ and $a_{2n}^{\! (2)}$ in general terms (thereby making it possible to consider momentum conservation in general terms), however, what \eqref{foldlag} becomes in specific cases of interest may be obscure. For this reason one representative of each of the families of folded theory is given here as an example - in each case the choice made is $\tilde{d} = d$ as it is this case that gives solitons non-trivial time delays.

\subsubsection{$c_2^{(1)}$ defect}

The case of $a_3^{\! (1)} \to c_2^{(1)}$ is of interest as it is the simplest canonical folding of the \ar theories. The starting point here is to take the unfolded Lagrangian \eqref{alag} with $u$ and $v$ as $a_3^{(1)}$ fields and $\chi$ similarly chosen to have three components\footnote{Generally in \ar it will be assumed that $\chi = \sum_{j=1}^r \chi_j \alpha_j$.}. The folding is achieved, as in section \ref{folding}, by setting
\begin{align*}
u_1 = u_3 = \frac{\phi_1}{2} \; , \; u_2 = \phi_2 \; , \; v_1 = v_3 = \frac{\psi_1}{2} \; , \; v_2 = \psi_2
\end{align*}
resulting in 
\begin{align}
\mathcal{L} &= \theta (-x) \mathcal{L}_{\phi} + \theta (x) \mathcal{L}_{\psi} \nonumber \\
& \quad + \delta (x) \left[ (\phi_1 - \psi_1)( \dot{\chi}_1 - \dot{\chi}_2 + \dot{\chi}_3 ) + (\phi_2 - \psi_2 )( -2 \dot{\chi}_1 + 2 \dot{\chi}_2 ) - D( \phi , \psi , \chi ) \right]   \label{c2lag}
\end{align}
where \small
\begin{align}
D( \phi , \psi , \chi ) &= d \left( e^{-\frac{\phi_1}{2} - \chi_3} + e^{\frac{\phi_1}{2} - \phi_2 + \chi_1} + e^{-\frac{\phi_1}{2} + \phi_2 - \chi_1 + \chi_2} + e^{\frac{\phi_1}{2} - \chi_2 + \chi_3} + (\phi \to \psi ) \right) \nonumber \\
& \quad + d^{-1} \left( e^{\frac{\phi_1}{2} + \chi_3} + e^{\frac{\phi_1}{2} - \chi_1} + e^{-\frac{\phi_1}{2} + \phi_2 + \chi_1 - \chi_2} + e^{\frac{\phi_1}{2} - \phi_2 + \chi_2 - \chi_3} + ( \phi \to \psi )  \right) \; . \label{c2pot} 
\end{align}
\normalsize
\\
The single algebraic constraint, $D_{\chi_1} + D_{\chi_2} = 0$, which can be seen to arise in the kinetic terms at the defect in \eqref{c2lag}, allows one of the degrees of freedom of the auxiliary field $\chi$ to be removed, leaving two degrees of freedom - the same number as have the $c_2^{(1)}$ fields $\phi$ and $\psi$.

\subsubsection{$d_3^{(2)}$ defect}

Note that $d_2^{(2)}$, possessing a single field, is just the sinh-Gordon case, so the first new case to consider here is $a_5^{\! (1)} \to d_3^{(2)}$. The Lagrangian \eqref{alag} should then be considered with $a_5^{\! (1)}$ fields and folding achieved by setting
\begin{align*}
u_1 = v_1 = 0  \; , \; u_2 = u_5 = \frac{\phi_1}{2} \; , \; u_3 = u_4 = \frac{\phi_2}{2} \; , \; v_2 = v_5 = \frac{\psi_1}{2} \; , \; v_3 = v_4 = \frac{\psi_2}{2}
\end{align*}
resulting in 
\begin{align}
\mathcal{L} &= \theta (-x) \mathcal{L}_{\phi} + \theta (x) \mathcal{L}_{\psi} \nonumber \\
& \quad + \delta (x) \left[ (\phi_1 - \psi_1)( - \dot{\chi}_1 + \dot{\chi}_2 - \dot{\chi}_4 + \dot{\chi}_5 ) + (\phi_2 - \psi_2 )( - \dot{\chi}_2 + \dot{\chi}_4 ) - D( \phi , \psi , \chi ) \right]   \label{d3lag}
\end{align}
where \small
\begin{align}
D( \phi , \psi , \chi ) &= d \left( 2 e^{-\chi_5} + 2 e^{-\chi_2 + \chi_3} \right) + d^{-1} \left( 2 e^{-\chi_1} + 2 e^{\chi_3 - \chi_4} \right) \nonumber \\
& \quad + d \left( e^{-\frac{\phi_1}{2} + \chi_1} + e^{\frac{\phi_1 - \phi_2}{2} - \chi_1 + \chi_2} + e^{\frac{-\phi_1 + \phi_2}{2} - \chi_3 + \chi_4} + e^{\frac{\phi_1}{2} - \chi_4 + \chi_5} + (\phi \to \psi ) \right) \nonumber \\
& \quad + d^{-1} \left( e^{-\frac{\phi_1}{2} + \chi_5} + e^{\frac{\phi_1}{2} + \chi_1 - \chi_2 } + e^{\frac{-\phi_1 + \phi_2}{2} + \chi_2 - \chi_3} + e^{\frac{\phi_1 - \phi_2}{2} + \chi_4 - \chi_5} + ( \phi \to \psi )  \right) \; . \label{d3pot} 
\end{align}
\normalsize
\\
The algebraic constraints in this case are $D_{\chi_1} + D_{\chi_5} = 0$, $D_{\chi_2} + D_{\chi_4} = 0$ and $D_{\chi_3}=0$, which may be used to reduce the number of degrees of freedom in $\chi$ from five down to two.

\subsubsection{$a_4^{\! (2)}$ defect}

Since a defect for $a_2^{\! (2)}$ is already known \cite{CZ09b,CZ10b}, the first new case arising from this analysis is $a_4^{\! (1)} \to a_4^{\! (2)}$. The folding is done by setting
\begin{align*}
u_1 = u_4 = \frac{\phi_1}{2} \; , \; u_2 = u_3 = \frac{\phi_2}{2} \; , \; v_1 = v_4 = \frac{\psi_1}{2} \; , \; v_2 = v_3 = \frac{\psi_2}{2}
\end{align*}
resulting in
\begin{align}
\mathcal{L} &= \theta (-x) \mathcal{L}_{\phi} + \theta (x) \mathcal{L}_{\psi} \nonumber \\
& \quad + \delta (x) \left[ (\phi_1 - \psi_1)( \dot{\chi}_1 - \dot{\chi}_3 + \dot{\chi}_4 ) + (\phi_2 - \psi_2 )( - \dot{\chi}_1 + \dot{\chi}_3 ) - D( \phi , \psi , \chi ) \right]   \label{a4lag}
\end{align}
where \small
\begin{align}
D( \phi , \psi , \chi ) &= d \left( 2 e^{-\chi_1 + \chi_2 } \right) + d^{-1} \left( 2 e^{\chi_2 - \chi_3} \right) \nonumber \\
& \quad + d \left( e^{-\frac{\phi_1}{2} - \chi_4 } + e^{\frac{\phi_1 - \phi_2}{2} + \chi_1 } + e^{\frac{-\phi_1 + \phi_2}{2} - \chi_2 + \chi_3} + e^{\frac{\phi_1}{2} - \chi_3 + \chi_3} + (\phi \to \psi ) \right) \nonumber \\
& \quad + d^{-1} \left( e^{-\frac{\phi_1}{2} + \chi_4} + e^{\frac{\phi_1}{2} - \chi_1 } + e^{\frac{-\phi_1 + \phi_2}{2} + \chi_1 - \chi_2} + e^{\frac{\phi_1 - \phi_2}{2} + \chi_3 - \chi_4} + ( \phi \to \psi )  \right) \; . \label{a4pot} 
\end{align}
\normalsize
\\
The algebraic constraints, $D_{\chi_1} + D_{\chi_3} = 0$ and $D_{\chi_2} = 0$, may be used to reduce the number of degrees of freedom of $\chi$ from four down to two.

\section{Discussion} \label{discus}

\subsection*{Conclusions}

The main conclusion of this paper is that certain \ar defects may be folded to give defects in the folded models of $c_n^{(1)}$, $d_n^{(2)}$ and $a_{2n}^{\! (2)}$ with Lagrangian description in the form \eqref{foldlag}. There are two strong reasons to believe that such defects are integrable.
\begin{itemize}
\item Firstly, energy and momentum conservation, when applied to a type I \ar defect, are in themselves enough to force the defect to be integrable \cite{CZ09a}. It is certainly plausible that the same holds for these folded defects.
\item Secondly, there is the rather simpler argument of classical scattering of solitons off the defect. This argument alone is suggestive of integrability as the solitons retain their form and so ought to conserve an infinite number of charges.
\end{itemize}
\noindent
Though not the main focus, this paper also furthers the work of \cite{KS} by using the relatively simple arena of \ar to construct solitons and breathers in $c_n^{(1)}$, $d_n^{(2)}$ and $a_{2n}^{\! (2)}$ - the breather solutions in particular have not been published before. \\
\\
One loose end regards the interpretation of the auxiliary field, $\chi$, in the folded defect. Since the algebraic constraints may be used to reduce the number of degrees of freedom of $\chi$ to that of a folded field, it seems highly likely that one may construct similar defects by starting in the folded model, i.e., without any reference to \ar. 

\subsection*{Future directions}

Note that all ATFT defects found thus far have been purely transmitting, agreeing with the findings of Delfino, Mussardo and Simonetti \cite{DMS94b}, so one clear extension of this work would be to find the quantum transmission matrices for the folded defects. The transmission matrices for type II defects in the \ar theories are found in \cite{CZ10b} and one would hope that in a similar way in which \cite{Khas} presents the folded bulk S-matrices in terms of the unfolded ones, the folded transmission matrices may be found. \\
\\
Another direction is to try to find defects for the other simply laced ATFTs ($d_s^{(1)}$, $e_6^{(1)}$, $e_7^{(1)}$ and $e_8^{(1)}$) which then might be folded such that all ATFTs are covered. In fact, there are possibly some implications already for $d_s^{(1)}$ defects in this paper due to the use of the non-canonical foldings $a_{2n-1}^{\! (1)} \to d_n^{(2)}$ and $a_{2n}^{\! (1)} \to a_{2n}^{\! (2)}$. For both of these folded theories the canonical way to fold is from a $d$ series ATFT rather than an $a$ series one (see figures \ref{ddfold} and \ref{dafold}), so there is the question of whether a $d_s^{(1)}$ defect ATFT, should such a thing exist, might be folded to give something of the form of \eqref{foldlag}. \\
\\
\begin{figure}
\centering
\begin{picture}(360,70)
 
\put(17,3){\circle{6}}
\put(17,39){\circle{6}}
\put(40,21){\circle{6}}
\put(66,21){\circle{6}}
\put(92,21){\circle{6}}
\put(118,21){\circle{6}}
\put(144,21){\circle{6}}
\put(167,3){\circle{6}}
\put(167,39){\circle{6}}

\put(43,21){\line(1,0){20}}
\put(69,21){\line(1,0){20}}
\put(95,21){\line(1,0){20}}
\put(121,21){\line(1,0){20}}

\put(144,24){\line(4,3){20}}
\put(144,18){\line(4,-3){20}}

\put(20,3){\line(4,3){20}}
\put(20,39){\line(4,-3){20}}

\put(60,45){\small{$d_{n+1}^{(1)}$}}

\put(175,21){\vector(1,0){20}}
\put(17,8){\vector(0,1){10}}
\put(17,34){\vector(0,-1){10}}
\put(167,8){\vector(0,1){10}}
\put(167,34){\vector(0,-1){10}}

\put(204,21){\circle*{6}}
\put(230,21){\circle{6}}
\put(256,21){\circle{6}}
\put(282,21){\circle{6}}
\put(308,21){\circle{6}}
\put(334,21){\circle{6}}
\put(360,21){\circle*{6}}

\put(206,22){\line(1,0){21}}
\put(206,20){\line(1,0){21}}

\put(233,21){\line(1,0){20}}
\put(259,21){\line(1,0){20}}
\put(285,21){\line(1,0){20}}
\put(311,21){\line(1,0){20}}

\put(358,22){\line(-1,0){21}}
\put(358,20){\line(-1,0){21}}

\put(260,38){\small{$d_{n}^{(2)}$}}

\end{picture}
\caption{$d_{n+1}^{(1)} \to d_{n}^{(2)}$.}
\label{ddfold}
\end{figure}
\begin{figure}
\centering
\begin{picture}(210,140)

\put(10,10){\circle{6}}
\put(33,28){\circle{6}}
\put(10,46){\circle{6}}
\put(56,46){\circle{6}}
\put(79,64){\circle{6}}
\put(56,82){\circle{6}}
\put(33,100){\circle{6}}
\put(10,118){\circle{6}}
\put(10,82){\circle{6}}

\put(13,10){\line(4,3){20}}
\put(36,28){\line(4,3){20}}
\put(33,31){\line(-4,3){20}}
\put(59,46){\line(4,3){20}}
\put(79,67){\line(-4,3){20}}
\put(56,85){\line(-4,3){20}}
\put(33,103){\line(-4,3){20}}
\put(33,97){\line(-4,-3){20}}

\put(10,15){\vector(0,1){20}}
\put(10,51){\vector(0,1){12}}
\put(10,113){\vector(0,-1){20}}
\put(10,79){\vector(0,-1){12}}
\put(43,41){\vector(0,1){20}}
\put(43,87){\vector(0,-1){20}}

\put(90,64){\vector(1,0){20}}

\put(60,113){\small{$d_{2n+2}^{(1)}$}}

\put(120,64){\circle*{3}}
\put(120,64){\circle{6}}
\put(146,64){\circle*{6}}
\put(172,64){\circle*{6}}
\put(198,64){\circle{6}}

\put(123,65){\line(1,0){21}}
\put(123,63){\line(1,0){21}}

\put(149,64){\line(1,0){20}}

\put(195,65){\line(-1,0){21}}
\put(195,63){\line(-1,0){21}}

\put(140,79){\small{$a_{2n}^{\! (2)}$}}

\end{picture}
\caption{$d_{2n+2}^{(1)} \to a_{2n}^{(2)}$.}
\label{dafold}
\end{figure}
\noindent
$\! \! \! \! \!$One can also look at the possibility of using defects to find more general integrable boundary conditions. The possibility of a boundary with an auxiliary field was considered in \cite{BD}; while the paper \cite{CZ12} considers a defect fused to a Ghoshal--Zamolodchikov type boundary \cite{GZ} in $a_1$.
Classically integrable boundary conditions, which happen to be highly restrictive (no free parameters), have been known for the folded theories $c_n^{(1)}$, $d_n^{(2)}$ and $a_{2n}^{\! (2)}$, for some time \cite{BCDR}. There is the possibility of fusing the defects here to such boundaries to give more general boundary conditions, though it is not immediately clear whether or not integrability would be preserved.

\section*{Acknowledgements}

The author wishes to thank Peter Bowcock and Ed Corrigan for their guidance and thought-provoking discussions. Thanks are also extended to the referee of a previous version of this paper. This work was supported by an STFC studentship. 

\begin{center}
\line(1,0){450}
\end{center}


\end{document}